%% file: MAPsurv_Final.tex
\documentclass[AMA,STIX1COL]{WileyNJD-v2}

\usepackage{moreverb}

\usepackage{graphics}
\usepackage{graphicx}
\usepackage{subfigure}
\usepackage{caption}
\usepackage{endfloat}
\usepackage{threeparttable}
\usepackage{enumitem}
\RequirePackage{fix-cm}

\newcommand\BibTeX{{\rmfamily B\kern-.05em \textsc{i\kern-.025em b}\kern-.08em
		T\kern-.1667em\lower.7ex\hbox{E}\kern-.125emX}}

\articletype{Article Type}%

\received{<day> <Month>, <year>}
\revised{<day> <Month>, <year>}
\accepted{<day> <Month>, <year>}

\begin{document}
	
	\title{{B}ayesian leveraging of historical control data for a clinical trial with time-to-event endpoint}
	
	\author[1]{Satrajit Roychoudhury*}
	
	\author[2]{Beat Neuenschwander}

	\authormark{Roychoudhury \textsc{et al}}

	\address[1]{\orgdiv{}, \orgname{Pfizer Inc}, \orgaddress{\state{New York, NY}, \country{USA}}}
	
	\address[2]{\orgdiv{}, \orgname{Novartis Pharma AG}, \orgaddress{\state{Basel}, \country{Switzerland}}}

	\corres{*Satrajit Roychoudhury, New York NY. \email{satrajit.roychoudhury@pfizer.com}}
	
	
	\abstract[Abstract]{The recent 21st Century Cures Act propagates innovations to accelerate the discovery, development, and delivery of 21st century cures. It includes the broader application of Bayesian statistics and the use of evidence from clinical expertise. An example of the latter is the use of trial-external (or historical) data, which promises more efficient or ethical trial designs. We propose a Bayesian meta-analytic approach to leveraging historical data for time-to-event endpoints, which are common in oncology and cardiovascular diseases. The approach is based on a robust hierarchical model for piecewise exponential data. It allows for various degrees of between trial-heterogeneity and for leveraging individual as well as aggregate data. An ovarian carcinoma trial and a non-small-cell cancer trial illustrate methodological and practical aspects of leveraging historical data for the analysis and design of time-to-event trials.}
	
		
	\maketitle
	

\section{Introduction}
\label{s:intro}

\input{./_sec1_intro}

\section{Methods}
\label{s:meth}

\input{./_sec2_meth}

\section{Application}
\label{s:app}
\input{./_sec3_app}

\section{Discussion}
\label{s:discuss}
\input{./_sec4_discuss}

\section*{Appendix A: Extraction of Data from Published Kaplan-Meier Plots}
\label{s:AppendixA}
\input{./_Appendix_A}
\section*{Appendix B: WinBUGS Code for MAC Analysis}
\label{s:AppendiB}
\input{./_Appendix_B}

\bibliography{bibfile}

\end{document}

%% file: _sec1_intro.tex
Historical data often provide valuable information for the design of a
new study. For example, sample size calculations depend on variability
and effect sizes from previous trials. However, current practice
usually ignores historical data in the analysis of a new trial. In
recent years, however, leveraging historical data in the analysis of
clinical trials has been encouraged by the European Medicines
Evaluation Agency (EMEA) \cite{EMEA2014}, the US Food and Drug
Administration (FDA) including the Prescription Drug User Fee Act VI
\cite{PDUFA6}, and the 21st Century Cures Act \cite{USact21}. Recent
examples in various phases of drug and medical device development
include French et al. \cite{French2012}, Hueber et
al. \cite{Hueber1693}, and Campbell \cite{Campbell2017}.

Leveraging historical data is appealing to practitioners and
regulators for reasons of improved efficiency (smaller and therefore
faster trials) and ethics (fewer patients assigned to a less effective
treatment). Yet this may be challenging and requires care. First,
identifying relevant historical data requires good judgment and
collaboration by the various stakeholders of a clinical trial. A
thorough review of the literature and other resources (e.g. registry
data) in a specific disease (e.g., study population, definition of
endpoint, data collection etc.) is important. This is ideally done
using systematic reviews techniques (Egger et al. \cite{egg1995srh},
Cochrane Collaboration \cite{CHMPmeta2001}, Rietbergen et
al. \cite{Rietbergen2011}), and involving a third party or independent
group may be useful. 

Second, a principled statistical approach, which leverages the
historical data while allowing for potential differences between
historical and actual data, is needed. Various statistical methods
have been proposed in this context: Pocock's bias model \cite{Poc76},
power priors (Ibrahim and Chen \cite{Ib00}), commensurate priors
(Hobbs et al. \cite{Hobbsetal2011}), and meta-analytic-predictive
priors (Spiegelhalter et al. \cite{Sp04}, Neuenschwander et
al. \cite{Neu10}, Schmidli et al. \cite{Schmidli2014}); for an
overview, see Viele et al. \cite{Viele2014} and Lewis et al. \cite{Lewis2019}.  They are very similar
and use hierarchical models that allow for different parameters in the
historical and actual trial, which will imply discounting of the
historical data when analyzing the new trial.

Here, we will be concerned with leveraging historical control data for
time-to-event endpoints (e.g., time to disease progression, time to
death). Such endpoints are the primary outcome in various therapeutic
areas, including oncology and cardiovascular diseases.  Among the
various methods we will use the \emph{meta-analytic-predictive (MAP)}
approach. This is essentially a Bayesian random-effects meta-analysis
of the historical data with the prediction of the parameter in the new
trial. The basic model, which assumes exchangeable parameters for the new and
 the historical trials, can be extended in various ways. We will extend the 
 \emph{MAP} methodology for one-dimensional parameters to piecewise exponential (PWE) time-to-event data. To hedge against potential prior-data conflict, we propose
 an extension of the basic exchangeability assumption by adding a robust mixture 
 component.

Although using historical data in clinical trials has gained
increasing interest recently, applications in the time-to-event
setting are sparse. A major challenge is the need of patient-level
data for most of the statistical time-to-event approaches (Murray et
al. \cite{Murray2014}, Bertsche et al. \cite{Bertsche2017}, and Hobbs
et al. 2013 \cite{Hobbs2013}). Our proposed methodology is applicable
for both patient-level and summary data (e.g. Kaplan Meier curves), the
latter often being available from publications.

The paper is structured as follows. In Section
\ref{s:meth} we introduce the basic \emph{MAP}
methodology and extend it to time-to-event data. Two applications are
discussed in Section \ref{s:app}, with emphasis on the analysis
of trial data in the presence of historical data, and the design of a
new trial leveraging historical data, respectively. Section
\ref{s:discuss} concludes with a discussion.

%% file: _sec2_meth.tex
In this section we discuss a meta-analytic-predictive (MAP) approach
to leveraging historical data for a time-to-event outcome. We consider
the randomized setting, with a control and test treatment for which
the use of historical data is confined to the control group. That is,
while the prior distribution for the control (baseline) hazard will be
informed by historical data, the prior for the treatment contrast
(proportional hazards parameter) will be weakly informative.

\subsection{The meta-analytic-predictive  (MAP) prior distribution}
The \emph{MAP} approach uses the historical control data from $J$
trials $Y_1,\ldots,Y_J$ to obtain the \emph{MAP} prior distribution
for the control parameter $\theta_{\star}$ in the new study
\begin{equation}
\label{methods::map1}
p(\theta_{\star} \vert Y_1,\ldots,Y_J)
\end{equation}
The derivation of (\ref{methods::map1}) is meta-analytic, using a
hierarchical model with trial-specific parameters
$\theta_{\star},\theta_1,\ldots,\theta_J.$
The simplest hierarchical model 
assumes exchangeable parameters across trials, which is usually
represented as 
\begin{equation}
\label{par::model1}
\theta_j = \mu + \epsilon_j, \quad \epsilon_j \sim N(0,\tau^2),
\quad j=\star,1,\ldots,J
\end{equation} 
While (\ref{par::model1}) allows for biases, they are assumed non-systematic. 
If needed, the basic model can be extended by replacing $\mu$ by
$X_j\beta$ to allow for systematic biases explained by covariates 
trial-specific covariates $X_j$.  

\subsection{A hierarchical model for piecewise exponential
 time-to-event data}

We assume piecewise exponential data and the time axis partitioned into $K$ intervals,
\begin{equation}
\label{methods:intervals}
 (I_{k-1},I_k], \quad k=1,...,K, \quad I_0=0
\end{equation}
For each historical trial and time interval, there are $n_{jk}$ control patients at risk (with a
corresponding total exposure time $E_{jk}$), of which $r_{jk}$
experience an event.
For each interval, the data model is Poisson
\begin{equation}
\label{methods::lik::histdata}
r_{jk} \vert \lambda_{jk}\sim \mbox{Poisson}(\lambda_{jk}E_{jk}), 
\quad j=1,\ldots,J; \, k=1,\ldots,K
\end{equation}
where $\lambda_{jk}$ is the hazard in interval $k$ of trial $j$.
Of interest is the prior for the control hazards $\lambda_{\star 1},
\ldots, \lambda_{\star K}$ in the new trial. 
The similarity of the new and the historical trials is captured in a
parameter model. The simplest model assumes normally distributed
log-hazard parameters $\theta_{jk} = \log(\lambda_{jk})$,
\begin{equation}
\label{methods::par::model}
\theta_{\star k},\theta_{1k},\ldots,\theta_{Jk} \vert \mu_k,\tau_k \sim N(\mu_k,\tau_k^2),
\quad k=1,\ldots,K
\end{equation}
For the across-trial mean parameters $\mu_k$, different implementations
are possible. First, ignoring the time structure, the parameters
may be assumed unrelated, with independent prior distributions
\begin{equation}
\label{methods:prior:mu}
\mu_k \sim  N(m_{\mu k},s_{\mu k}^2), \quad k=1,\ldots,K
\end{equation}
Second, the time structure of $\mu_1,\ldots,\mu_K$ may be modelled
using, for example, a multivariate normal distribution with a
structured covariance matrix (e.g. AR(1)), or a dynamic linear model
\begin{eqnarray}
\label{methods:dynamic:mu}
\mu_k & \sim  & N(\eta_k,\sigma_k^2), \quad k=1,\ldots,K \\
\eta_k & = & \eta_{k-1} + \rho_{k-1} \quad k=2,\ldots,K
\end{eqnarray}
The latter will be used in the applications of Section \ref{s:app},
which gives more details for the prior distributions. 

Finally, prior distribution for the across-trial standard deviations
$\tau_k$ on the log-hazard scale are required. We will use half-normal
priors
\begin{equation}
\label{methods:prior:tau}
\tau_k  \sim  \mbox{half-normal}(s_{\tau k})
\end{equation}
with scale parameter $s_{\tau k}$ representing anticipated heterogeneity. A weakly-informative half-Normal prior with $s_{\tau k}$=0.5 puts approximately 5\% probability for values of $\tau_{k}$ greater than 1. This allows small to large between trial heterogeneity. As for the mean parameters $\mu_k$, rather than assuming unrelated $\tau_k$ parameters, a multivariate normal distribution or dynamic linear model could be used
for the $\log(\tau)$ parameters.

From the above data model, parameter model, and prior distributions, 
the \emph{MAP} prior for the vector of log-hazards in control group
then follows as the conditional distribution of
$\theta_{\star_1},\ldots,\theta_{\star_K}$
given historical data,
\begin{equation}
\label{methods::map2}
p(\theta_{\star_1} \ldots, \theta_{\star_K} \vert r,E)
\end{equation}
where $r$ and $E$ denote the number of events and exposure times
across all historical trials and time intervals. The \emph{MAP} prior
can be obtained via MCMC. WinBUGS/JAGS code is given in the Appendix B.

\subsection{Prior effective number of events}
\label{sec::ene}
When desiging a new trial, knowing the amount of information
introduced by the \emph{MAP} prior is useful. It is sometimes
expressed as the prior \emph{effective sample size (ESS)}, or, in the
time-to-event setting, the \emph{effective number of events
	(ENE)}. Various methods have been suggested (Malec \cite{Malec2001},
Morita et al. \cite{Morita2008}, Neuenschwander et al. \cite{Neu10}, Pennello and Thompson \cite{Pennello2008}).
They compare the information of the prior (variance or precision) to
the one from one observation (or event). While conceptually similar,
they can lead to suprisingly different \emph{ESS} or \emph{ENE}.

We will use an improvement, the \emph{expected
local-information-ratio $ESS_{ELIR}$} as introduced by Neuenschwander et al. \cite{Neuenschwander2019}. It is defined as 
the expected (under the prior) ratio of prior information
$i(p(\theta))$ to Fisher information $i_F(\theta)$
\begin{equation}
ESS_{ELIR} = E_{\theta} \left\{ \frac{i(p(\theta))}{i_F(\theta)} \right\}
\end{equation}
where $i(p(\theta)) = -d^2 \log p(\theta) / d\theta^2$ and 
$i_F(\theta) = -E_{Y_1 \vert \theta} \lbrace d^2 \log p(Y_1 \vert
\theta) / d\theta^2 \rbrace$.

It can be shown that, like the other methods, $ESS_{ELIR}$ fulfills
the necessary condition of being consistent with the well-known
\emph{ESS} for conjugate one-parameter exponential families. Unlike
previous methods, however, it also fulfills a basic predictive
criterion. That is, for a sample of size $N$, the expected posterior
\emph{ESS} (under the prior distribution) is the sum of the prior
\emph{ESS} and $N$.

For the piecewise exponential model with log-hazard parameters
$\theta_{\star k}$  $(k=1,\ldots,K)$ in the new trial, the Fisher
information $i_F(\theta_{\star k})$ for one event is $1$. Since the
\emph{MAP} priors for the log-hazard parameters are available only as
an MCMC sample, approximations for $p(\theta_{\star k})$ will be
needed. Mixtures of standard distributions are convenient and can
approximate these distributions with any degree of accuracy
(Dalal and Hall \cite{dallal1983}, Diaconis and Ylvisaker \cite{dia1984qpo}). 
Fitting mixture distributions can be done by various procedures (e.g.,
SAS \cite{SASref}, R package RBesT \cite{web2019rbest},\cite{Weberetal2019}).  Finally,
after having obtained the effective number of events for each
log-hazard parameter, the total effective number of events of the \emph{MAP}
prior follows as $ENE = \sum_{k=1}^K ESS_{ELIR}(\theta_{\star k})$.

\subsection{Analysis of the data in the new trial}
\label{sec::mapmac}
We now turn to the analysis of the new, randomized trial, 
where the control treatment $(C)$ will be compared to a test
treatment $(T)$. For the piecewise exponential model, the control and
treatment data in interval $k$ follow Poisson distributions 
\begin{equation}
\label{methods::lik::newdata}
r_{C\star k} \vert \lambda_{C \star k} \sim
\mbox{Poisson}(\lambda_{\star k} E_{C \star k}), \quad
r_{T\star k} \vert \lambda_{T \star k} \sim 
\mbox{Poisson}(\beta \lambda_{\star _k} E_{T\star k})
\end{equation}
where $\lambda_{\star k} = \exp(\theta_{\star k})$ are the control hazards, and
$\beta$ is the Cox proportional hazards parameter. Prior information
for the control log-hazards $\theta_{\star k}$
are given by the \emph{MAP} prior (\ref{methods::map2}), whereas the prior
for $\beta$ will usually be weakly-informative.

When leveraging historical data, the analysis for the new trial can be
done in two ways. 
\begin{itemize}
\item Following the \emph{MAP} approach, one would combine the
  \emph{MAP} prior (\ref{methods::map2}) from historical controls with
  the likelihood of the new data (\ref{methods::lik::newdata}). This
  is complicated because the \emph{MAP} prior for the $K$ control
  log-hazards $\theta_{\star}$ is not known analytically but only
  available as an MCMC sample. Approximations of the \emph{MAP} prior
  could be used; for example, by matching a multivariate normal
  distribution with the same means, standard deviations, and
  correlations; or, mixture approximations to the \emph{MAP} priors
  could be used. The latter would allow for accurate approximations
  but would also be technically more complex.
\item Alternatively, the \emph{meta-analytic-combined (MAC)} approach
  can be used. It consists of the hierarchical analysis of
  all the data and has been shown to be equivalent (Schmidli et
  al. \cite{Schmidli2014}) to the \emph{MAP} approach. Importantly, it
  is computationally easier than the two-step \emph{MAP} approach and
  is therefore used here. For applications and respective code,
  see Section \ref{s:app} and the Appendix, respectively. 
\end{itemize}

\subsection{Robust meta-analyses}
\label{sec::robustmap}
To address the possibility of prior-data conflict, Schmidli et
al. \cite{Schmidli2014} proposed robust versions of \emph{MAP} or
\emph{MAC} analyses. The idea builds on exending the standard
parameter model (exchangeable parameters), allowing the control
parameter in the new trial to be non-exchangeable with the historical
parameters. The implementation uses a mixture distribution with
weights $w$ and $1-w$ for exchangeability and non-exchangeablity,
respectively.

This idea can be easily extended to the time-to-event setting as follows:
for each interval $k$ the robust version assumes exchangeablity,
$\theta_{\star k} \sim N(\mu_k,\tau_k^2)$, with probability $w_k$ and
non-exchangeability with probability $1-w_k$. For the latter,
interval-specific prior distributions 
\begin{equation}
\label{methods::par::robust}
\theta_{\star k} \sim
N(m_{\theta k},s_{\theta k}^2)
\end{equation}
are needed. To achieve good robustness, weakly-informative priors
should be used (for example, approximate unit-information priors,
i.e., $s_k=1$ for exponential data on the log-scale).  Of note, 
even though the $w_k$ are fixed, the Bayesian calculus ensures dynamic 
updating. That is, the prior weights $w_k$ will be updated to posterior
weights depending on the similarity of the new and historical data.
Eventually on may be interested in extending this idea to all trials by 
introducing trial-specific mixture weights $w_{jk}$, $j=1,\ldots,J$, $k=1,\ldots,K$.

%% file: _sec3_app.tex
We now discuss two applications of leveraging historical time-to-event
data. The first illustrates the joint analysis of historical and new
data with the meta-analytic-combined approach of
Section \ref{sec::mapmac}, whereas the second emphasizes the design
of a new trial in the presence of historical data.

\subsection{Analysis with historical data}
\label{s:anal}
\begin{figure}[htp]
     \vspace*{0.5in}
     \centering
     \includegraphics[scale=0.45]{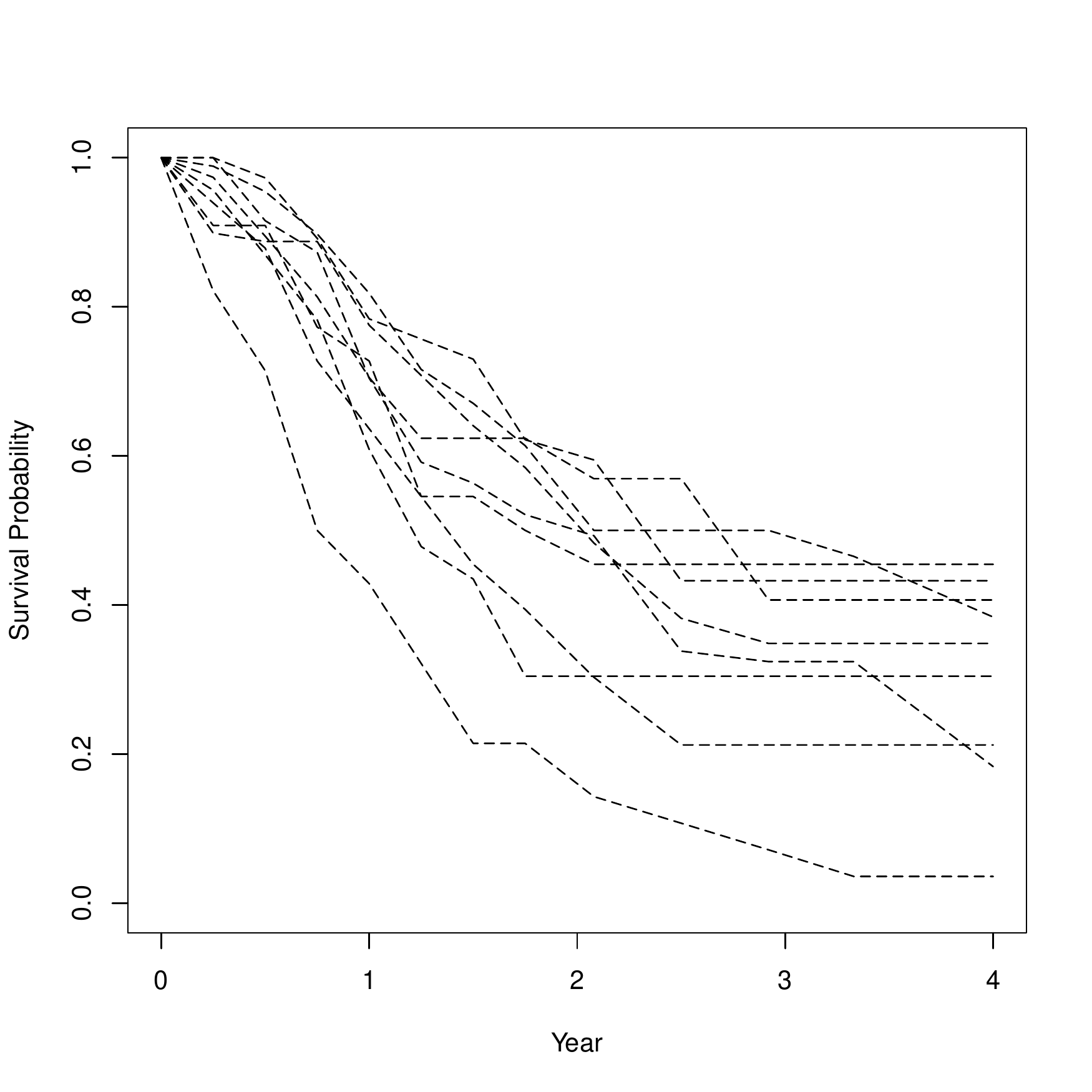}
      \includegraphics[scale=0.45]{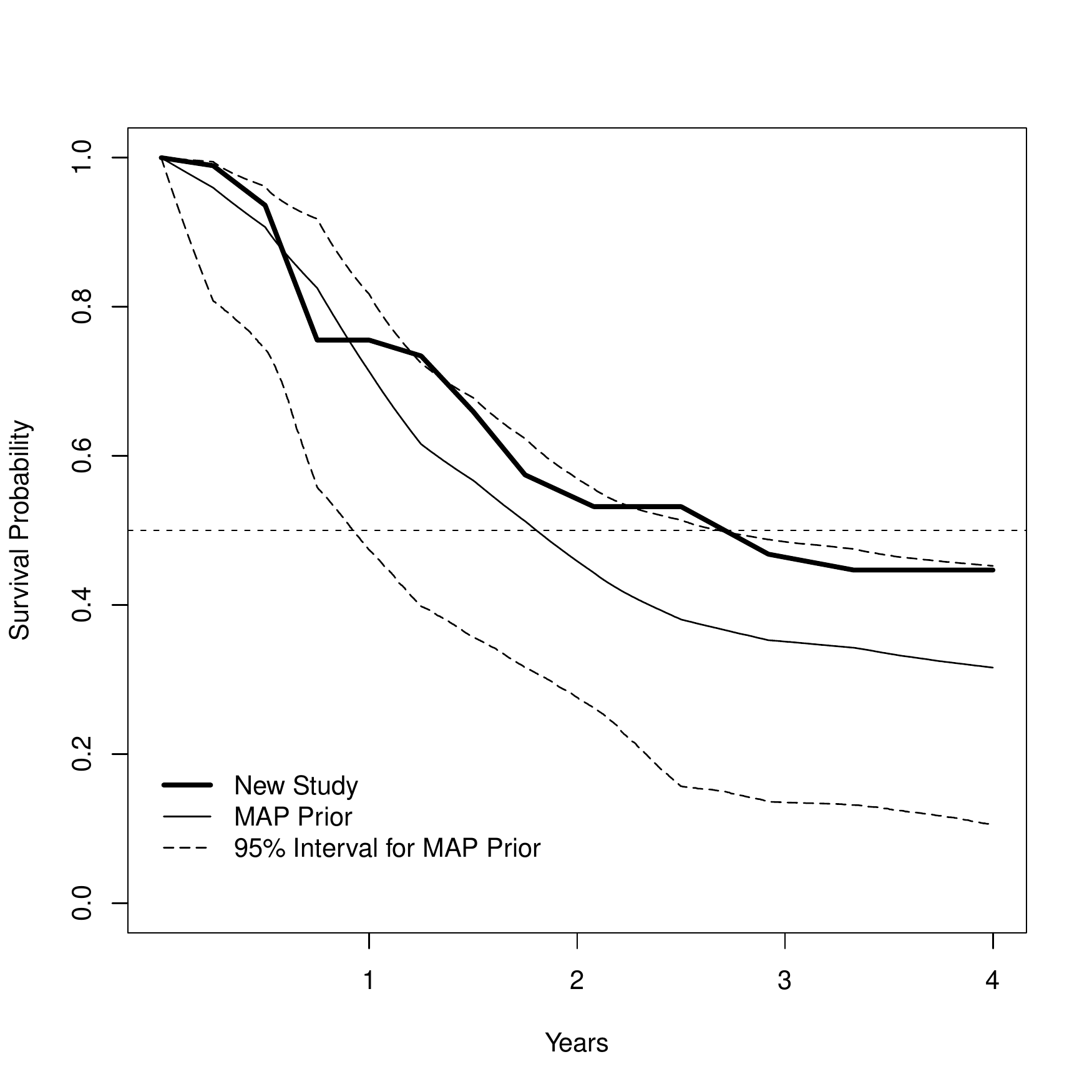}
      \caption{Application 1: Kaplan-Meier curves for the control group in nine historical
        trials (left panel), meta-analytic-predictive (MAP) prior
        (median (solid line), 95\%-interval (dashed line) in right
        panel), and  Kaplan-Meier curve for new data (thick solid line,
        right panel)}
     \label{fig:KM_FIOCCO}
\end{figure}

Voest et al. \cite{Voe89} and Fiocco et al.  \cite{Fio09SiM}
investigated survival data from ten studies on patients with advanced
epithelial ovarian carcinoma. For our purpose, we assume the last
study in Table \ref{tab:Fioccodata} as the study of interest and the
data of the remaining nine studies as the historical data. Thus, we
want to infer the survival curve of the last study while leveraging
the data from the other studies.  The Kaplan-Meier plots in Figure
\ref{fig:KM_FIOCCO} show considerable heterogeneity across the nine
historical trials (left panel), with median survivals ranging from 1
to 2.9 years.

Fiocco et al. \cite{Fio09SiM} used the following 12 invervals (in years): [0-0.25], (0.25-0.50],
(0.50-0.75], (0.75-1], (1-1.25], (1.25-1.50],
(1.50-1.75], (1.75- 2.08], (2.08-2.50], (2.50-2.92], (2.92-3.33], and (3.33-4].
For the following meta-analytic analyses, the number of deaths and
total exposure time per interval have been extracted from the
published Kaplan-Meier plots using the Parmar et al. \cite{Parmar1998}
approach: the
total exposure time ($E_{jk}$) for interval $(I_{k-1},I_k]$ and trial
$j$ is calculated as
\begin{eqnarray*} E_{jk} = \frac{L_{jk}}{2}\times (r_{jk} + c_{jk}) +
L_{jk}\times (n_{jk} - r_{jk} - c_{jk}); \qquad j= 1,\ldots,10, k=1,\ldots,12
\end{eqnarray*}
Here, $L_{jk}$ is the interval length, and $n_{jk}$, $r_{jk}$ and
$c_{jk}$ are the number of patients at risk, dead, and censored,
respectively. The data extraction assumes constant censoring rates for
each interval and all deaths at mid-interval times. Table
\ref{tab:Fioccodata} summarizes the data for the 12 intervals in the
ten studies.

\begin{table}[ht]
\caption{Application 1: number of deaths/exposure time for 12
  intervals in ten studies} \label{tab:Fioccodata}
\smallskip 
   \begin{threeparttable}
\begin{tabular}{ccccccccccc}
  \hline
          & \multicolumn{9}{c}{historical studies}\\  
  \cline{2-10}
interval  &   1 &  2 & 3 & 4 & 5 & 6 & 7 & 8 & 9 & new study \\
(years)   &     &    &   &   &   &   &   &   &   & \\
\hline
0.00-0.25 & 1/9.4 & 9/21.1  & 1/21.9 & 1/5.6 & 5/6.4 & 0/17.8 & 2/8.0 & 0/9.2 & 2/5.2 & 1/23.4 \\
0.25-0.50 & 3/8.8 & 1/19.9  & 3/21.4 & 2/5.2 & 3/5.4 & 6/17.0 & 2/7.5 & 1/9.1 & 0/5.0 & 5/22.6 \\
0.50-0.75 & 3/7.9 & 0/19.8  & 5/20.4 & 2/4.8 & 6/4.2 & 3/15.9 & 5/6.6 & 3/8.6 & 3/4.6 & 17/19.9 \\
0.75-1.00 & 4/7.0 & 10/18.5 & 7/18.9 & 4/4.0 & 2/3.2 & 12/14.0 & 3/5.6 & 4/7.8 & 1/4.1 & 0/17.8 \\
1.00-1.25 & 3/6.1 & 6/16.5  & 9/16.9 & 3/3.1 & 3/2.6 & 8/11.5 & 3/4.9 & 1/7.1 & 4/3.5 & 2/17.5 \\
1.25-1.50 & 0/5.8 & 6/15.0  & 4/15.2 & 1/2.6 & 3/1.9 & 2/10.2 & 3/4.1 & 1/6.9 & 0/3.0 & 7/16.4 \\
1.50-1.75 & 0/5.8 & 5/13.6  & 5/14.1 & 3/2.1 & 0/1.5 & 3/9.6 & 2/3.5 & 4/6.2 & 1/2.9 & 8/14.5 \\
1.75-2.08 & 2/7.3 & 9/15.7  & 10/16.2& 0/2.3 & 2/1.7 & 2/11.9 & 3/3.8 & 1/7.4 & 1/3.5 & 4/17.2 \\
2.08-2.50 & 0/8.8 & 9/16.2  & 0/18.5 & 0/2.9 & 1/1.5 & 11/12.4 & 3/3.6 & 6/8.0 & 0/4.2 & 0/21.0 \\
2.50-2.92 & 6/7.6 & 3/13.6  & 0/18.3 & 0/2.9 & 1/1.0 & 1/9.9 & 0/2.9 & 0/6.7 & 0/4.2 & 6/19.7 \\
2.92-3.33 & 0/6.2 & 0/12.5  & 3/17.0 & 0/2.9 & 1/0.6 & 0/9.4 & 0/2.9 & 0/6.6 & 0/4.1 & 2/17.4 \\
3.33-4.00 & 0/10.0 & 0/20.1  & 7/24.5 & 0/4.7 & 0/0.7 & 10/12.1 & 0/4.7 & 0/10.7 & 0/6.7 & 0/27.5\\
\hline
\end{tabular}
   \end{threeparttable}
\end{table}

The prior information about the hazards in study ten is captured by the
\emph{MAP} prior. The respective prior for survival (median and
95\%-intervals in right panel of Figure \ref{fig:KM_FIOCCO} ) shows a
median of approximately 1.8 years (95\% interval 0.9 to 2.7 years). The
prior effective number of events (Section 2.3) is 58. Figure ~\ref{fig:KM_FIOCCO} 
shows the KM plots of nine historical data (left panel) and MAP prior (right panel).
The KM curve of the tenth trial (thick solid line in the right panel) is also shown
 in Figure ~\ref{fig:KM_FIOCCO}.

After having access to the data from trial 10, for illustration we
will compare two \emph{MAC} analyses and the stratified analysis:
\begin{enumerate}[label=(\roman*)]
\item \emph{EX}: full exchangeability \emph{MAC} analysis.  This is
  model (\ref{methods::par::model}) with between-trial standard
  deviations $\tau_{k}$ following half-normal priors with scale 0.5,
  which cover small to large heterogeneity (95\% interval
  (0.02,1.12)). For $\eta_{1}$, unit-information prior $N(-1.171,
  1)$, centered at log(0.31) (the overall estimated log-hazard for
  death), is used.  $N(0,1)$ priors are used for $\rho _{i}$, $i=
  1\ldots k$.
\item \emph{EXNEX}: exchangeability-nonexchangeability \emph{MAC}
  analysis. This is the robust analysis of Section
  \ref{sec::robustmap}, where for the tenth trials we assume weight 0.5 for
  exchangeability \emph{\emph{EX}} (\ref{methods::par::model}) and 0.5
  for nonexchangeability \emph{(NEX)}
  (\ref{methods::par::robust}). For remaining nine trials full exchangeability is assumed (weight=1).
  For \emph{EX}, the prior assumptions
  in (i) are used, whereas for \emph{NEX} the priors are
  weakly-informative with means centered at the mean of the MAP prior
  and variances approximately worth one observation.
\item \emph{STRAT}: this is the analysis of the data from the tenth study only,
  ignoring the data from the historical trials. For $\eta_{1}$, a non-informative prior $N(0,10^2)$ is used. Similar to EX model, $N(0,1)$ priors are used for $\rho _{i}$, $i=
  1\ldots k$.
\end{enumerate}

For the three analyses, the posterior summaries for yearly survival
rates in trial ten are summarized in Table \ref{tab:Fiocco_MAC_Sum}.
The \emph{EX} analysis provides smaller survival rates compared to
the stratified analysis, and median survival is 2.01 years, considerably
different from the observed median (2.7 years). On the other hand, the
\emph{EXNEX} analysis is more robust, with a median survival of 2.59 months.

\begin{figure}[htp]
     \vspace*{0.5in}
     \centering
     \includegraphics[scale=1.0]{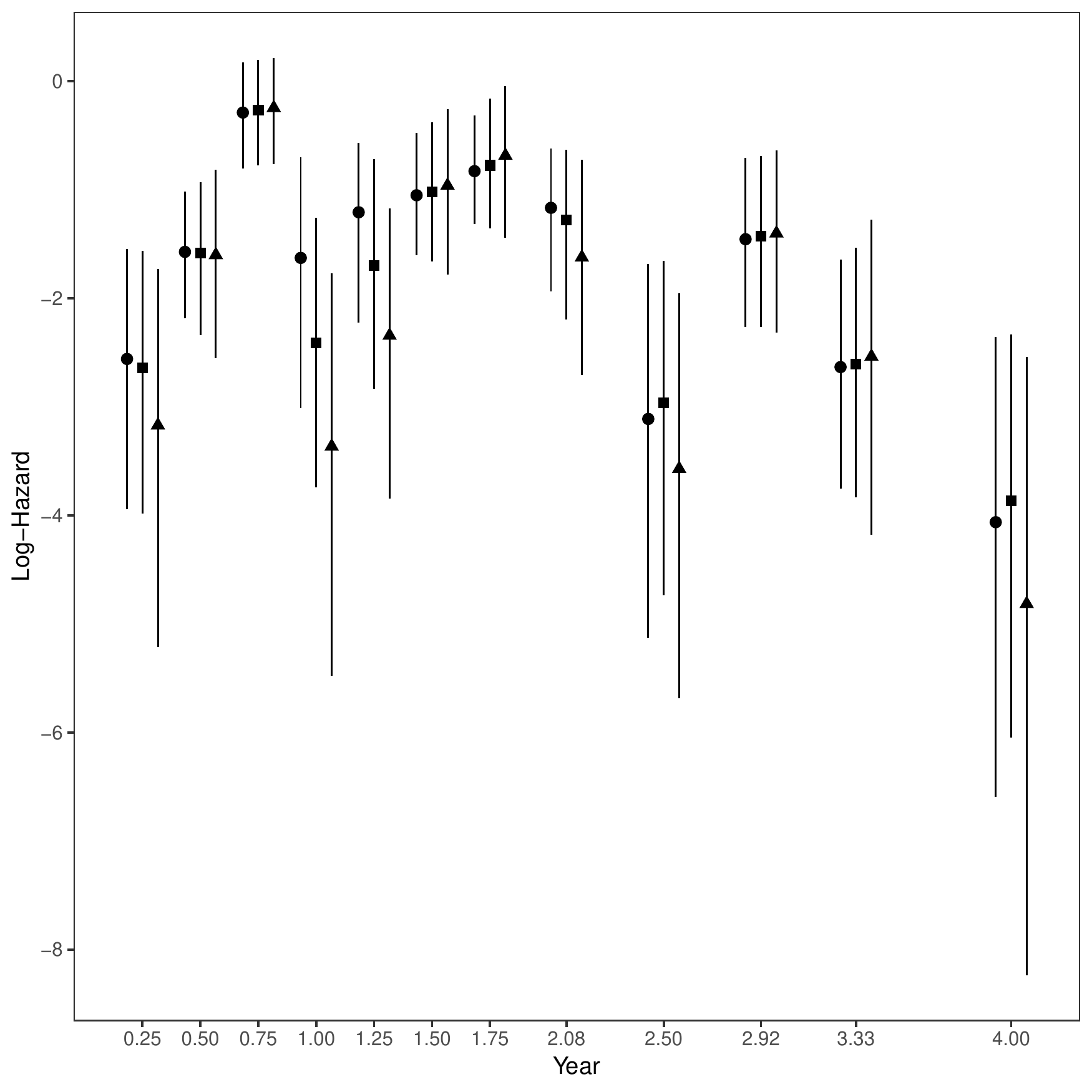}
     \caption{Application 1: posterior medians and 95-\% intervals of log-hazard for 12
       intervals for EX (circle), EXNEX (square) and stratified
       analysis (triangle) in new study}
     \label{fig:FIOCCO_MAC}
\end{figure}

\begin{table}[ht]
  \caption{Application 1: posterior median (95\%-interval) for yearly
    survival rates and median survival, and
    two approximations of the  posterior effective number of events (ENE) for
    exchangeability, exchangeability-nonexchangeability, and stratified analysis} 
\label{tab:Fiocco_MAC_Sum}
\centering
\smallskip 
\begin{tabular}{cccc}
\hline
                            &       EX             &    EXNEX           & STRAT  \\
survival rate               &                      &                    &            \\
 1 year                     &   0.72 (0.57, 0.86)  &  0.74 (0.59, 0.86) &  0.75 (0.60, 0.87)\\
 2 year                     &   0.50 (0.31, 0.71)  &  0.53 (0.31, 0.75) &  0.54 (0.32, 0.77) \\
 3 year                     &   0.43 (0.22, 0.67)  &  0.45 (0.22, 0.71) &  0.47 (0.23, 0.73) \\
 4 year                     &   0.41 (0.19, 0.67)  &  0.44 (0.20, 0.71) &  0.44 (0.20, 0.71) \\
median survival (years)     &   2.24 (1.59, 3.19)  &  2.62 (1.68, 5.05) &  7.90 (1.69, 39.86) \\
\hline
\end{tabular}
\end{table}

\subsection{Study design with historical data}
\label{s:degn}
The second application is concerned with the design of a randomized
phase II trial for lung cancer with historical control data. As of
2018, lung cancer is the most common cause of cancer-related death in
men and women, responsible for 1.76 million deaths annually
worldwide ( WHO Cancer Factsheet 2018 \cite{WHOcancer2018}).

The aim of the phase II proof-of-concept \emph{(POC)} study was to
compare a new treatment \emph{(T)} against an active control treatment
\emph{(C)} for patients with locally advanced recurrent or metastatic
lung cancer. The primary endpoint was progression-free survival
\emph{(PFS)}, defined as the time from treatment assignment to disease
progression or death from any cause. Superiority of the treatment
\emph{T} is typically established by a statistically significant
log-rank test or an upper confidence limit of the Cox proportional
hazard-ratio (HR) less than 1.

For the control treatment, the median PFS for this population is
approximately five months. Assuming a 45\% reduction in \emph{PFS}
$(HR=0.55)$ for the new treatment, a 2:1 (T:C) randomization, an
enrollment rate of 30 patients per month, a one-sided 2.5\% level of
significance and 90\% power, the study would require 133 events (230
patients). Due to limited resources, however, it was decided to
leverage historical data and enroll only 130
patients. For this modified design, the final analysis was planned
after 110 events.

\begin{figure}[htp]
     \vspace*{0.5in}
     \centering
     \includegraphics[scale=1.0]{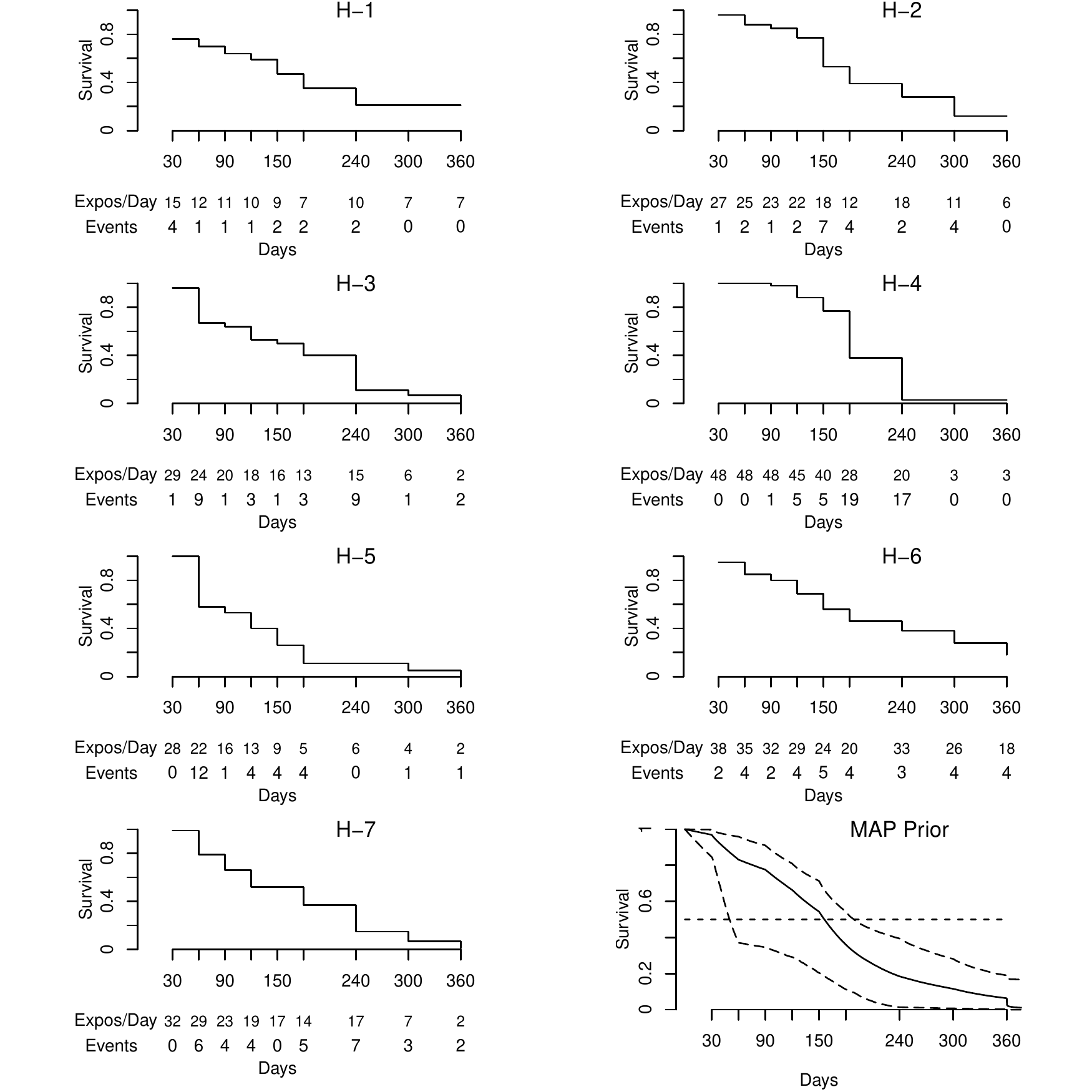}
     \caption{Application 2: Kaplan-Meier curves for seven historical studies and
       meta-analytic-predictive (MAP) prior (medians and 95\%-intervals) for
       new study}
     \label{MAC_Sim_Histdata}
\end{figure}

After an extensive search, seven historical studies with data for the
active control were identified by the clinical team and disease area
experts. Only Kaplan-Meier plots were available from the
literature. Figure \ref{MAC_Sim_Histdata} shows the Kaplan-Meier
plots as well as the number of events and exposure time for the
following intervals (in days): (0-30], (30-60], (60-90], (90-120],
(120-150], (150-180], (180-240], (240-300], and (300-360]. The data
extraction from the Kaplan-Meier plots was done by the method of
similar to Parmar et al. \cite{Parmar1998} ( for details see Appendix A). The Figure \ref{MAC_Sim_Histdata} shows the rather heterogeneous historical studies (medians vary between 3.5 months to 6 months) and the MAP prior (median= 4.88 months; 95\% interval (1.6 months, 6.2 months)). The prior effective  number of events (Section 2.3) is 36. 

The frequentist operating characteristics (type-I error, power, bias,
and root-mean-square error of the log-hazard ratio) of the design were
assessed for exponential data, two scenarios for the HR (1 and 0.5),
and various scenarios for the control median (3.5, 4.5, 5, 5.5,
6.5, and 7.5 months). Study success was defined by the Bayesian criterion
$P(\mbox{HR}<1 \vert \mbox{data}) >0.975$ or equivalently $P( \beta < 0 \vert \mbox{data}) >0.975$. For 2000 simulated trials, the
above metrics were assessed for four models:
\begin{enumerate}[label=(\roman*)]
\item \emph{EX}: full exchangeability \emph{MAC} analysis.  This is
  model (\ref{methods::par::model}) with between-trial standard
  deviations $\tau_{k}$ following half-normal priors with scale 0.5,
  which cover small to large heterogeneity (95\% interval
  (0.02,1.12)). For $\eta_{1}$, a unit-information prior $N(-5.167,
  1)$, centered at log(0.00575) (the overall estimated log-hazard for
  death), was used.  $N(0,1)$ priors were used for $\rho _{i}$, $i=
  1\ldots k$.
\item \emph{EXNEX90}: exchangeability-nonexchangeability \emph{MAC}
  analysis. This is the robust analysis, with weights 0.9 for
  exchangeability \emph{\emph{EX}} (\ref{methods::par::model}) and 0.1
  for nonexchangeability \emph{(NEX)}
  (\ref{methods::par::robust}). For \emph{EX}, the prior assumptions
  in (i) were used, whereas for \emph{NEX} the priors were assumed as
  weakly-informative with means centered at the mean of the MAP prior
  and variances equals 1  (worth one observation).
\item \emph{EXNEX50}: same as \emph{EXNEX90} but with 50-50 weights
  for exchangeability and nonexchangeability.
\item \emph{STRAT}: the analysis of the data from the new study only,
  ignoring the data from the historical trials and
  assuming $N(0,10^2)$ priors for the log-hazards in each interval. 
\end{enumerate}

\begin{table}[h]
	\caption{Application 2: type-I error and power (\%) for different
          control medians (3.5 to 7.5 months)  and treatment effects
          (hazard ratio = 1 or 0.55) for EX, EXNEX90, EXNEX50, and
          STRAT analysis}
\label{OCrslt}
	\begin{center}
	\begin{threeparttable}	
		\begin{tabular}{ccccccc} \hline
scenarios & control  &  treatment   &       EX  &    EXNEX90 &   EXNEX50 & STRAT\\
          & median   &  median      & 	        &            &		     &     \\
\cline{2-7}      
          & \multicolumn{6}{c}{HR=1 (type-I error)}\\
\cline{2-7}
 1        & 3.5      &   3.5     & $<$ 0.01 &  $<$ 0.01  &  $<$ 0.01 & 2.6\\    
 2        & 4.5      &   4.5     &      3.7 &       2.7  &      2.1  & 2.5\\
 3       & 5.0      &   5.0     &      5.3 &       5.1  &      4.0  & 2.5\\
 4       & 5.5      &   5.5     &      9.0 &       8.3  &      5.9  & 2.8 \\
 5        & 6.5      &   6.5     &     16.3 &      14.5  &      9.4  & 2.7\\
 6        & 7.5      &   7.5     &     23.7 &      20.9  &     13.2  &  2.6\\       
\cline{2-7}
          & \multicolumn{6}{c}{HR=0.55 (power)}\\
\cline{2-7}
 7        & 3.5      &   6.4     &  91.7    &      90.9  &     88.6  & 85.2 \\    
 8        & 4.5      &   8.2     &  97.3    &      96.9  &     94.5  & 84.9\\
 9       & 5.0      &   9.1     &  97.0    &      96.8  &     96.1  & 84.6\\
10       & 5.5      &   10.0    &  98.9    &      98.7  &     97.1  & 85.0 \\
11        & 6.5      &   11.8    &  99.3    &      99.2  &     98.0  & 84.8\\
12        & 7.5      &   13.6    &  99.7    &      99.5  &     98.5  & 84.5\\       			
			\hline
		\end{tabular}
\end{threeparttable}
	\end{center}
\end{table}

\begin{figure}[htp]
	\vspace*{0.5in}
	\centering
	\includegraphics[scale=1.0]{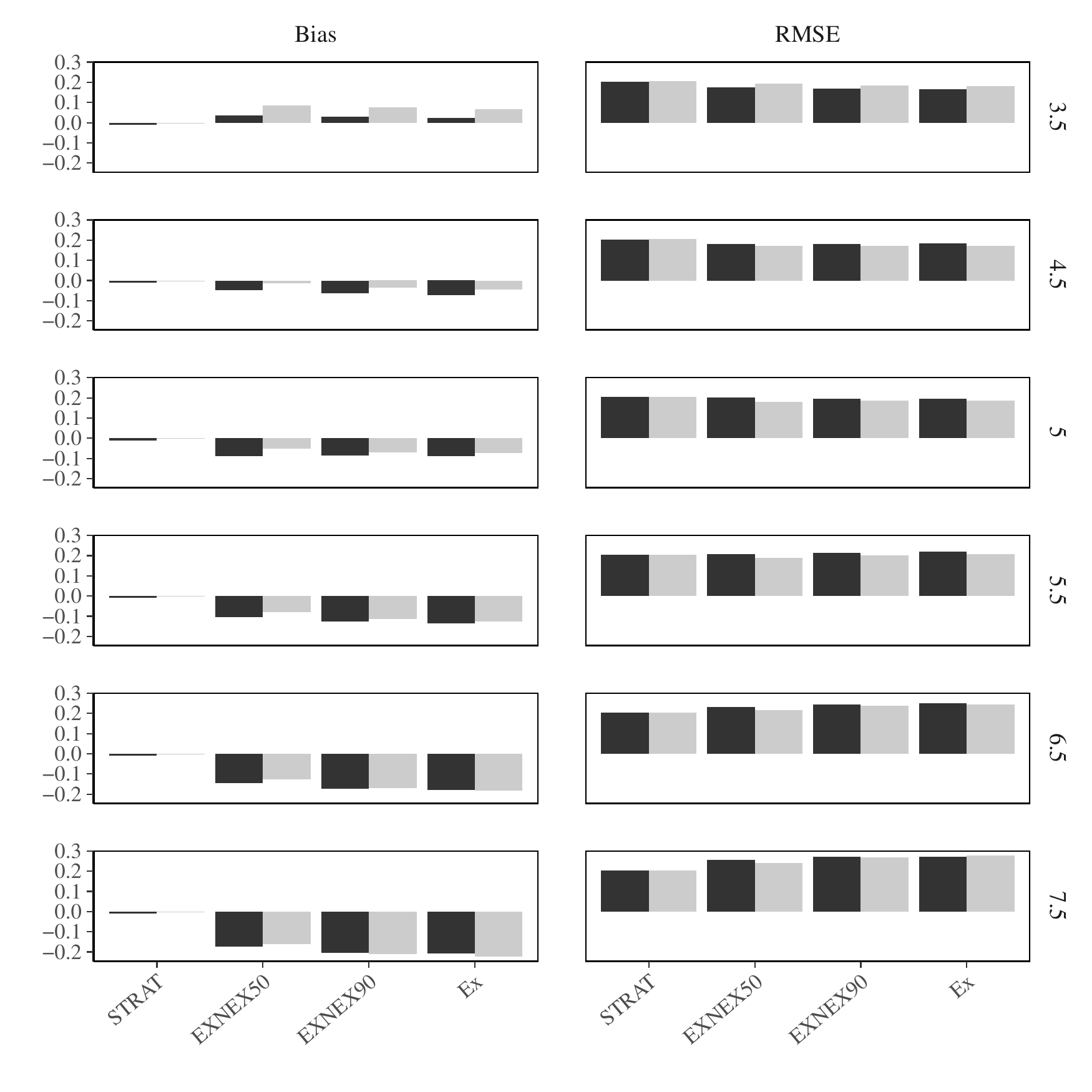}
	\caption{Application 2: bias (left panel) and root-mean-square error (RMSE)
          (right panel) of log(HR) for different control medians and
          hazard-ratios for STRAT, EXNEX50, EXNEX90, and EX analysis;
          black and gray bars show results for HR=1 and HR=0.55,
          respectively. }
	\label{OCrslt2}
\end{figure}

Table \label{OCrslt} and Figure	\ref{OCrslt2} summarize the operating
characteristics for the four models.Table \label{OCrslt} shows the type-I error and power for different scenarios. For the type-I error (HR=1), \emph{EX} and \emph{EXNEX90} show fairly similar results. If the assumed control median is
considerably smaller (3.5 months) than the one from the \emph{MAP} prior ($\approx 5$ months, Figure \ref{MAC_Sim_Histdata}), the type-I error is much smaller than 2.5\%.  On the other hand, if the control median is much larger (6.5 or 7.5 months), the type-I error is larger than 10\%. For the more robust \emph{EXNEX50} model, however, the type-I error is much less affected. Finally, for the Bayesian stratified analysis, the type-I error is close to the expected 2.5\%. For power (HR=0.55), the analyses that leverage the historical data exhibit substantial gains compared
to the stratified analyses, and power gains increase with increasing true control medians. 

The bias on the log-HR scale (left panel of Figure \ref{OCrslt2})
increases with increased leveraging (from stratified to fully      
exchangeable) and true control medians much smaller (3.5 months) or
much larger (7.5 months) than suggested by the historical data (5
months). For the non-robust \emph{EX} model, the bias is
approximately --20\% for the worst case scenario (true control median
6.5 months) but is generally much smaller for the robust analyses and
true control medians closer to the expected 5 months.The root-mean-square error (on log-HR scale) (Figure \ref{OCrslt2}) shows gain in efficiency in EX, EXNEX50, EXNEX90 when the new trial data are aligned with historical data.

%% file: _sec4_discuss.tex
Clinical trials results are usually interpreted in the context of
other relevant data. In addition to such informal considerations of
historical data, or more generally co-data (Neuenschwander et
al. \cite{Neuenschwander2016}), we have considered leveraging
historical data in the Bayesian analysis of a new clinical trial with
a time-to-event endpoint.  Historical control data are increasingly
used in earlier stages of drug development. As these studies primarily
inform company-internal decisions, the advantages are often considered
to outweigh potential risks. In phase III clinical trials, however,
historical controls are currently rare. Yet the regulatory environment
is evolving in special areas such as rare diseases, pediatric
populations, and non-inferiority trials (EMEA
\cite{ema2006gct,ema2001gci,ema2006}).

Leveraging historical control data has advantages. First, and most
importantly, it allows randomizing fewer patients to the control
group. This will shorten the duration of a clinical trial and hence
lead to faster decisions. It will also decrease trial costs, as fewer
patients will be needed. Second, if the control is ineffective
(e.g., for placebo), historical data designs may also be more ethical
because fewer patients receive the ineffective treatment
(Berry \cite{berry2004}). 

The selection of historical data requires care and should follow
recommendations for systematic reviews. This minimizes the risk of
systematic biases, which can arise as a result of, for example,
changes in standard of care over time, differences in
inclusion/exclusion criteria, confounding environmental factors, or
the evolution of diagnostic tools. The robust mixture approach
proposed here mitigates some of these problems but cannot compensate
for using biased data.

When analyzing the data of the new trial, a suitable model that allows
for different degrees of similarity between historical and new data,
is needed. Hierarchical models are the most obvious choice. For
piecewise exponential time-to-event data, we have discussed a flexible
model with interval-specific exchangeable hazard parameters and
between-trial standard deviations.  For few trials, which is typical
for most settings, information on between-trial variability will
usually be sparse. Yet valuable information may be available from
similar disease settings, which may lead to more informative prior
distributions for the between-trial standard deviations (Turner et al.
\cite{tur2012peh,tur2015pdb}).  In the applications of Section
\ref{s:app}, due to lack of additional information, we have used 
prior distributions that cover the range of small to large
heterogeneity. 

Widening the scope to real-world evidence, leveraging more but
possibly less relevant data may be of interest. This will require
adjustments to the methodology of Section \ref{s:meth} with
regard to bias and heterogeneity. Having access to relevant predictors
that explain anticipated biases will be key, and including these
predictors (via meta-regression or propensity score methods) will be
needed.  Moreover, if historical data are of different quality, this
may be accounted for by using different between-trial standard
deviations.  Further research will be needed to incorporate such
extensions in the time-to-event setting.

%% file: _Appendix_A.tex
For the seven historical trials in Application 2 (NSCLC example),we could only get the Kaplan-Meier plots from the published articles. The data extraction (number of ... ) was done using the following steps:

\begin{enumerate}
\item The time axis is divided into intervals of 30 days up-to 180 days and 60 days afterwards. Based on historical data the maximum follow up time is restricted to 360 days.
\item Probabilities of remaining progression-free at pre-specified time points are extracted using Engauge Digitizer  \cite{Engauge2019}, a freeware that converts an image to numbers.
\item The number of patients at risk in a specific interval is calculated by multiplying the total number of patients with the probability of remaining progression-free at the beginning of the interval.
\item The number of events (progression or death) within a time interval are approximated by using the probabilities of remaining progression-free in the current and next interval as follows; 
 \begin{eqnarray*}
\log(\hat{S(t_{i+1})}) -  \log(\hat{S(t_{i})}) =  1 - \frac{d_{i}}{n_{i}}
 \end{eqnarray*}
 where $d_{i}$ is the number of events and $n_{i}$ is the risk set for interval $[t_{i}, t_{i+1})$
 
\item The exposure time for an interval is the interval length if the patient did not have an event or was not censored. Otherwise, the exposure of the patient is half the interval length. The total exposure time for an interval is the sum of the individual exposure times at the risk set.
\end{enumerate}

%% file: _Appendix_B.tex
\subsection*{B.1 Main WinBUGS Code}

\begin{verbatim}

##Required R packages

library(MASS)
library(coda)
library(R2WinBUGS)
library(survival)
library(RBesT)

##Main WinBUGS code 

MAC.Surv.WB <- function()
{
# tau.study: Between study variation (Half normal)

Prior.tau.study.prec <- pow(Prior.tau.study[2],-2)

for ( t in 1:Nint) {

tau.study[t] ~ dnorm(Prior.tau.study[1], Prior.tau.study.prec) %_% I(0, )
tau.study.prec[t] <- pow(tau.study[t],-2)
}

#tau.time: correlation for piecewise exponential pieces (log-normal)

Prior.tau.time.prec <- pow(Prior.tau.time[2],-2)
tau.time ~ dlnorm(Prior.tau.time[1], Prior.tau.time.prec)
tau.time.prec <- pow(tau.time,-2)

# priors for regression parameter: h covariates

for (h in 1:Ncov) {
prior.beta.prec[h] <- pow(Prior.beta[2,h],-2)
beta[h,1] ~ dnorm(Prior.beta[1,h],prior.beta.prec[h])

}

###########################
#EXNEX structure of mu
###########################

#EX structure for mu is 1st order NDLM 

mu.prec.ex <- pow(Prior.mu.mean.ex[2],-2)
mu.mean.ex ~ dnorm(Prior.mu.mean.ex[1],mu.prec.ex)

mu1.ex ~ dnorm(mu.mean.ex,tau.time.prec)
mu.ex[1] <- mu1.ex

prec.rho.ex <- pow(Prior.rho.ex[2],-2)

for(t in 2:Nint){

mu.dlm.ex[t] <- mu.ex[t-1] + rho.ex[t-1]

#variance of mu: discount factor X tau.time.prec 

mu.time.prec.ex[t] <- tau.time.prec/w.ex

mu.ex[t] ~ dnorm(mu.dlm.ex[t], mu.time.prec.ex[t])

rho.ex[t-1] ~ dnorm(Prior.rho.ex[1], prec.rho.ex)

}

w.ex ~dunif(w1,w2)

#NEX structure for mu: unrelated structure 
for( s in 1:Nstudies){ 
for(t in 1:Nint) {

prior.mu.prec.nex[s,t] <- pow(prior.mu.sd.nex[s,t],-2)
mu.nex[s,t] ~ dnorm(prior.mu.mean.nex[s,t],prior.mu.prec.nex[s,t])

}
}

# hazard-base: for each study (+ population mean + prediction) and time-interval t, no covariates
for ( s in 1:Nstudies) {

for ( t in 1:Nint) {
RE[s,t] ~ dnorm(0,tau.study.prec[t])

Z[s,t] ~ dbin(p.exch[s,t],1)

#For ex

log.hazard.base.ex[s,t] <- mu.ex[t] + step(Nstudies-1.5)*RE[s,t]

#For Nexch

log.hazard.base.nex[s,t] <- mu.nex[s,t]

log.hazard.base[s,t] <- Z[s,t]*log.hazard.base.ex[s,t]+(1-Z[s,t])*log.hazard.base.nex[s,t]

hazard.base[s,t] <- exp(log.hazard.base[s,t])
}
}

# likelihood: pick hazards according to study and time-invervals (int.low to int.high) for each 
#observation j
# note: hazard is per unit time, not depending on length of interval t
for(j in 1:Nobs) {
for ( t in 1:Nint) {
# log.hazard for all time intervals
log(hazard.obs[j,t]) <- log.hazard.base[study[j],t] + inprod(X[j,1:Ncov],beta[1:Ncov,1])
}

#Poisson likelihood        

alpha[j] <- (inprod(hazard.obs[j,int.low[j]:int.high[j]],int.length[int.low[j]:int.high[j]])
                                       /sum(int.length[int.low[j]:int.high[j]]))*exp.time[j]
n.events[j] ~ dpois(alpha[j])
n.events.pred[j] ~ dpois(alpha[j])

}

# outputs of interest (from covariate patterns in Xout)
for ( h in 1:Nout) {
for ( s in 1:Nstudies) {
# mean (index = Nstudies)
# surv: pattern x study x time
surv1[ h,s,1] <- 1
for ( t in 1:Nint) {
log(hazard[h,s,t]) <- log.hazard.base[s,t] + inprod(Xout[h,1:Ncov],beta[1:Ncov,1])
log.hazard[h,s,t] <- log(hazard.base[s,t])
surv1[h,s,t+1] <- surv1[h,s,t]*exp(-hazard[h,s,t]*int.length[t])
surv[h,s,t] <- surv1[h,s,t+1]
}
}
}

######### Prediction##########

for ( h in 1:Nout) {
for ( t in 1:Nint) {
hazard.pred[h,t] <- hazard[h,Nstudies,t]
log.hazard.pred[h,t] <- log(hazard[h,Nstudies,t]+pow(10,-6))
surv.pred[h,t] <- surv[h,Nstudies,t]
}
}

for (j in 1:Ncov) {
beta.ge.cutoffs[j] <- step(beta[j,1]-beta.cutoffs[j,1])
}

}  

\end{verbatim}

\subsection*{B.2 R wrapper function for running the main WinBUGS code}

\begin{verbatim}

##Description of arguments of the function

#Nobs              =  Total number of data points
#study             =  Study indicator
#Nstudies          =  Number of studies
#Nint              =  Number of intervals
#Ncov              =  Number of covariates
#int.low, int.high =  Interval indicator
#int.length        =  Length of interval
#n.events          =  Number of events at each interval
#exp.time          =  Exposure time for each interval
#X                 =  Important covariates
#Prior.mu.mean.ex  =  Mean and standard deviation for normal priors for exchangeability for the first interval
#Prior.rho.ex      =  Mean and standard deviation for normal priors for random gradient of 1st order NDLM
#w1, w2            =  Upper and lower bound for uniform prior of the smoothing factor for 1st order NDLM model for exchangeability 
#prior.mu.mean.nex =  Mean of normal prior for non-exchangeability
#prior.mu.sd.nex   =  Standard deviation for normal prior of non-exchangeability
#p.exch            =  Prior probability of exchangeability
#Prior.beta        =  Mean and standard deviation for normal priors of regression coefficients  
#beta.cutoffs      =  Cut-off for treatment effect
#Prior.tau.study   =  Scale parameter of half-normal prior for between trial heterogeneity
#Prior.tau.time    =  Mean and standard deviation for log-normal prior for variance compoment of 1st order NDLM
#MAP.Prior         =  If TRUE: derives MAP prior 
#pars              =  Parameters to keep in each MCMC run
#bugs.directory    =  Directory path where WinBUGS14.exe file resides
#R.seed            =  Seed to generate initial value in R (requires for reprducibility) 
#bugs.seed         =  WinBUGS seed (requires for reprducibility)

MAC.Surv.anal <- function(Nobs  = NULL,
                          study             = NULL,
                          Nstudies          = NULL,
                          Nint              = NULL,
                          Ncov              = 1,
                          Nout              = 1,
                          int.low           = NULL,
                          int.high          = NULL,
                          int.length        = NULL,
                          n.events          = NULL,
                          exp.time          = NULL,
                          X                 = NULL,
                          Xout              = matrix(0,1,1),
                          Prior.mu.mean.ex  = NULL, 
                          Prior.rho.ex      = NULL,
                          w1                = 0,
                          w2                = 1,
                          prior.mu.mean.nex = NULL,
                          prior.mu.sd.nex   = NULL,
                          p.exch            = NULL,
                          Prior.beta        = NULL, 
                          beta.cutoffs      = NULL,
                          Prior.tau.study   = NULL, 
                          Prior.tau.time    = NULL,
                          MAP.prior         = FALSE,
                          pars              = c("tau.study","log.hazard","mu.ex", "hazard"),
                          bugs.directory    = "C:/Users/roychs04/Documents/Folder/Software/WinBUGS14",
                          R.seed            = 10,
                          bugs.seed         = 12
                         )
{  

set.seed(R.seed)

beta.cutoffs <- cbind(NULL,beta.cutoffs)

if(MAP.prior){Nstudies <- Nstudies+1}

#WinBUGS model

model <- MAC.Surv.WB

#Data for JAGS format

data     = list("Nstudies", "Nint", "Ncov", "Nobs", "Nout","study","int.low","int.high","int.length",
"n.events","exp.time",
"X","Xout",
"Prior.mu.mean.ex","Prior.rho.ex",
"prior.mu.mean.nex","prior.mu.sd.nex",
"p.exch","w1","w2",
"Prior.beta",
"beta.cutoffs",
"Prior.tau.study","Prior.tau.time"
)

#Initial values

hazard0 = (sum(n.events)+0.5)/sum(exp.time)

initsfun = function(i)
list(
mu1.ex = rnorm(1,log(hazard0),0.25),
tau.study = rgamma(12,1,1),
tau.time  = rgamma(1,1,1),
mu.mean.ex = rnorm(1,log(hazard0),0.1),
rho.ex = rnorm(Nint-1,0,0.05),
w.ex = runif(1,0,1),
beta=cbind(NULL,rnorm(Ncov,0,1))
)

inits <- lapply(rep(1,3),initsfun)

#WinBUGS run

fit = bugs(
data=data,
inits=inits,
par=pars,
model=model,
n.chains=3,n.burnin=8000,n.iter=16000,n.thin=1,
bugs.directory= bugs.directory,
bugs.seed= bugs.seed,
DIC= TRUE,
debug= FALSE
)

fit$sims.matrix = NULL
fit$sims.array = NULL

#WinBUGS summary

summary <- fit$summary 
R2WB <- fit

output <- list(summary=summary,R2WB =R2WB)

return(output)

}

\end{verbatim}

\subsection*{B.3 FIOCCO Analysis}

\begin{verbatim}

##FIOCCO data set

FIOCCO.n.events <- c(1,  3,  3,  4,  3,  0,  0,  2,  0,  6,  0,  0,  9,  1,  0, 10,  6,  6,  5,  9,
                     9,  3,  0,  0,  1,  3,  5,  7,  9,  4,  5, 10,  0,  0,  3,  7,  1,  2,  2,  4, 
                     3,  1,  3,  0,  0,  0,  0,  0,  5,  3,  6,  2,  3,  3,  0,  2,  1,  1,  1,  0, 
                     0,  6,  3, 12,  8,  2, 3,  2, 11,  1,  0, 10,  2,  2,  5,  3,  3,  3,  2,  3, 
                     3,  0,  0,  0,  0,  1,  3,  4, 1,  1,  4,  1,  6,  0,  0,  0,  2,  0,  3,  1, 
                     4,  0,  1,  1,  0,  0,  0,  0,  1,  5, 17, 0,  2,  7,  8,  4,  0,  6,  2,  0)

FIOCCO.exp.time <- c(  9.4,  8.8,  7.9,  7.0,  6.1,  5.8,  5.8,  7.3,  8.8,  7.6,  6.2, 10.0, 21.1, 
                      19.9, 19.8, 18.5, 16.5, 15.0, 13.6, 15.7, 16.2, 13.6, 12.5, 20.1, 21.9, 21.4, 
                      20.4, 18.9, 16.9, 15.2, 14.1, 16.2, 18.5, 18.3, 17.0, 24.5,  5.6,  5.2,  4.8, 
                       4.0,  3.1,  2.6,  2.1,  2.3,  2.9,   2.9,  2.9,  4.7,  6.4,  5.4,  4.2,  3.2, 
                        2.6,  1.9,  1.5,  1.7, 1.5,  1.0,  0.6,  0.7, 17.8, 17.0, 15.9, 14.0, 11.5, 
                      10.2,  9.6, 11.9, 12.4,  9.9,  9.4, 12.1,  8.0,  7.5,  6.6,  5.6,  4.9,  4.1, 
                       3.5,  3.8,  3.6,  2.9,  2.9,  4.7,  9.2,  9.1,  8.6,  7.8,  7.1,  6.9,  6.2,  
                       7.4,  8.0,  6.7,  6.6, 10.7,  5.2,  5.0,  4.6,  4.1,  3.5,  3.0,  2.9,  3.5, 
                       4.2,  4.2,  4.1,  6.7, 23.4, 22.6, 19.9, 17.8, 17.5, 16.4, 14.5, 17.2, 21.0, 
                      19.7, 17.4, 27.5)
\end{verbatim}

\subsubsection*{B.3.1  Derivation of MAP Prior using First 9 Studies}

\begin{verbatim}
FIOCCO.MAP.Prior <- MAC.Surv.anal(Nobs              = 108,
                                  study             = sort(rep(1:9,12)),
                                  Nstudies          = 9,
                                  Nint              = 12,
                                  Ncov              = 1,
                                  Nout              = 1,
                                  int.low           = rep(1:12,9),
                                  int.high          = rep(1:12,9),
                                  int.length        = c(0.25, 0.25, 0.25, 0.25, 0.25, 0.25, 0.25,
                                                        0.33, 0.42, 0.42, 0.41, 0.67),
                                  n.events          = FIOCCO.n.events[1:108],
                                  exp.time          = FIOCCO.exp.time[1:108],
                                  X                 = matrix(0,108,1),
                                  Prior.mu.mean.ex  = c(0, 10), 
                                  Prior.rho.ex      = c(0,10),
                                  prior.mu.mean.nex = matrix(rep(0,108), nrow=9, ncol=12),
                                  prior.mu.sd.nex   = matrix(rep(1,108), nrow=9, ncol=12),
                                  p.exch            = matrix(rep(1,108), nrow=9),
                                  Prior.beta        = matrix(c(0,10), nrow=2) , 
                                  beta.cutoffs      = 0,
                                  Prior.tau.study   = c(0, 0.5), 
                                  Prior.tau.time    = c(-1.386294, 0.707293),
                                  MAP.prior         = TRUE, 
                                  pars              = c("tau.study","log.hazard.pred"),
                                  R.seed            = 10,
                                  bugs.seed         = 12
)

print(FIOCCO.MAP.Prior$summary)

#Calculation of ENE

FIOCCO.MAP.ess <- NULL

for(l in 1:12){

prior.ss.int <- FIOCCO.MAP.Prior$R2WB$sims.list$log.hazard.pred[,,l]
prior.ss.int.mix <- RBesT::automixfit(prior.ss.int, type="norm")

FIOCCO.MAP.ess[l] <- RBesT::ess(prior.ss.int.mix, sigma=1)

}

#Prior ESS (mixture)

FIOCCO.ESS.nevent <- sum(FIOCCO.MAP.ess)

\end{verbatim}

\subsubsection*{B.3.1  MAC analysis of FIOCCO data }

\begin{verbatim}

FIOCCO.anal.1 <- MAC.Surv.anal(Nobs              = 120,
                               study             = sort(rep(1:10,12)),
                               Nstudies          = 10,
                               Nint              = 12,
                               Ncov              = 1,
                               Nout              = 1,
                               int.low           = rep(1:12,10),
                               int.high          = rep(1:12,10),
                               int.length        = c(0.25, 0.25, 0.25, 0.25, 0.25, 0.25, 0.25, 
                                                     0.33, 0.42, 0.42, 0.41, 0.67),
                               n.events          = FIOCCO.n.events,
                               exp.time          = FIOCCO.exp.time,
                               X                 = matrix(0,120,1),
                               Prior.mu.mean.ex  = c(-1.1711, 1), 
                               Prior.rho.ex      = c(0,1),
                               prior.mu.mean.nex = matrix(rep(0,120), nrow=10, ncol=12),
                               prior.mu.sd.nex   = matrix(rep(1,120), nrow=10, ncol=12),
                               p.exch            = matrix(rep(1,120), nrow=10),
                               Prior.beta        = matrix(c(0,10), nrow=2) , 
                               beta.cutoffs      = 0,
                               Prior.tau.study   = c(0, 0.5), 
                               Prior.tau.time    = c(-1.386294, 0.707293),
                               R.seed            = 10,
                               bugs.seed         = 12
                               )

print(FIOCCO.anal.1$summary)

\end{verbatim}

\subsubsection*{B.3.2  Robust MAC analysis of FIOCCO data}

\begin{verbatim}

FIOCCO.anal.2 <- MAC.Surv.anal(Nobs              = 120,
                               study             = sort(rep(1:10,12)),
                               Nstudies          = 10,
                               Nint              = 12,
                               Ncov              = 1,
                               Nout              = 1,
                               int.low           = rep(1:12,10),
                               int.high          = rep(1:12,10),
                               int.length        = c(0.25, 0.25, 0.25, 0.25, 0.25, 0.25, 0.25, 0.33,
                                                     0.42, 0.42, 0.41, 0.67),
                               n.events          = FIOCCO.n.events,
                               exp.time          = FIOCCO.exp.time,
                               X                 = matrix(0,120,1),
                               Prior.mu.mean.ex  = c(-1.1711, 1), 
                               Prior.rho.ex      = c(0,1),
                               prior.mu.mean.nex = matrix(rep(c(-1.8625303, -1.6057708, -1.1242566,  
                                                                -0.5940037, -0.5921193, -1.2484085, 
                                                                -1.0011891, -0.9291769, -1.3337843, 
                                                                -2.1254918, -2.9740698, -2.7570149), 
                                                                10), 
                                                                nrow=10, byrow = T),
                              prior.mu.sd.nex   = matrix(rep(1,12*10), nrow=10, byrow=T),
                              p.exch            = rbind(matrix(rep(1,108), nrow=9), rep(0.5, 12)),
                              Prior.beta        = matrix(c(0,10), nrow=2) , 
                              beta.cutoffs      = 0,
                              Prior.tau.study   = c(0, 0.5), 
                              Prior.tau.time    = c(-1.386294, 0.707293),
                              R.seed            = 10,
                              bugs.seed         = 12
                              )

print(FIOCCO.anal.2$summary)

\end{verbatim}